\documentclass[prb,preprint,aps,superscriptaddress,longbibliography]{revtex4-1}
\usepackage[]{color}

\usepackage{graphicx, epstopdf}
\usepackage{subfigure}
\usepackage{float}
\usepackage{amsmath}
\usepackage{amsfonts}
\usepackage{amssymb}
\usepackage{bm}
\usepackage{subfigure}
\usepackage{caption}
\usepackage{hyperref}
\begin{document}

\title{Predicting Dirac semimetals based on Sodium Ternary Compounds}

\author{Bo Peng}
\thanks{These two authors contributed equally to this work.}
\affiliation{Beijing National Laboratory for Condensed Matter Physics,
and Institute of Physics, Chinese Academy of Sciences, Beijing 100190, China}
\affiliation{Department of Optical Science and Engineering and Key Laboratory of Micro and Nano Photonic Structures (Ministry of Education), Fudan University, Shanghai 200433, China.}

\author{Changming Yue}
\thanks{These two authors contributed equally to this work.}
\affiliation{Beijing National Laboratory for Condensed Matter Physics,
  and Institute of Physics, Chinese Academy of Sciences, Beijing
  100190, China}
\affiliation{University of Chinese Academy of Sciences, Beijing 100049, China}

\author{Hao Zhang}
\email{zhangh@fudan.edu.cn}
\affiliation{Department of Optical Science and Engineering and Key Laboratory of Micro and Nano Photonic Structures (Ministry of Education), Fudan University, Shanghai 200433, China.}

\author{Zhong Fang}
\affiliation{Beijing National Laboratory for Condensed Matter Physics,
  and Institute of Physics, Chinese Academy of Sciences, Beijing
  100190, China}
\affiliation{Collaborative Innovation Center of Quantum Matter,
  Beijing, China}

\author{Hongming Weng}
\email{hmweng@iphy.ac.cn}
\affiliation{Beijing National Laboratory for Condensed Matter Physics,
  and Institute of Physics, Chinese Academy of Sciences, Beijing
  100190, China}
\affiliation{Collaborative Innovation Center of Quantum Matter,
  Beijing, China}


\date{\today}

\begin{abstract}
Predicting new Dirac semimetals, as well as other topological materials, is challenging since the relationship between crystal structure, atoms and band topology is complex and elusive. Here, we demonstrate an approach to design Dirac semimetals via exploring chemical degree of freedom. Based on understanding of the well-known Dirac semimetal, Na$_3$Bi, three compounds in one family, namely Na$_2$MgSn, Na$_2$MgPb and Na$_2$CdSn, are located. Furthermore, hybrid-functional calculations with improved accuracy for estimation of band inversion show that Na$_2$MgPb and Na$_2$CdSn have the band topology of Dirac semimetals. The nontrivial surface states with Fermi arcs on the (100) and (010) surfaces are shown to connect the projection of bulk Dirac nodes. Most importantly, the candidate compounds are dynamically stable and have been experimentally synthesized. The ideas in this work could stimulate further predictions of topological materials based on understanding of existing ones.
\end{abstract}

\flushbottom
\maketitle

\thispagestyle{empty}

\section{Introduction}


Dirac semimetals (DSMs)~\cite{Young2012,Wang2012a,Wang2013,Weng2016,Armitage2018} are the 3D analogues of graphene~\cite{Novoselov2004} with and only with Dirac nodes on the Fermi level. These Dirac nodes are formed by band crossing, and the low-energy excitation around them leads to quasiparticles described by Dirac equation as emergent massless Dirac fermions.~\cite{Liu2014b, Armitage2018,Bradlyn2016,Bernevig2018,Orlita2014,Yang2014a} Up to now, there have been three classes of DSM proposed. One is the Dirac nodes with four-fold essential degeneracy, which is enforced by the nonsymmorphic symmetry at the high-symmetric momenta on the boundary of the Brillouin zone.~\cite{Young2012} The second is the accidental degenerate Dirac nodes, which appears as the topological phase transition critical point between different topological insulating states~\cite{Murakami2007}. The third one is also an accidental DSM, but the band crossing points are caused by band inversion and protected by proper crystal symmetry.~\cite{Wang2012a,Yang2014a} DSMs serve as a singular point of various topological states, such as topological insulators, Weyl semimetals, nodal line semimetals and triple-point semimetals~\cite{Weng2017}. DSMs exhibit many novel properties, such as high carrier mobility~\cite{Zdanowicz1975}, unique surface states with Fermi arcs~\cite{Wang2012a,Wan2011} and negative longitudinal magnetoresistivity due to the chiral anomaly.~\cite{Xiong2015,Gorbar2014}

The breakthrough in the search for stable DSMs~\cite{Yang2014a} is achieved in the series of studies on Na$_3$Bi~\cite{Wang2012a,Liu2014b} and Cd$_3$As$_2$~\cite{Wang2013,Liu2014c,Borisenko2014,Jeon2014,Neupane2014}, both of which were first proposed through first-principles calculations. They present good examples of the realization of the DSM in the above third class. The Dirac nodes are induced by band inversion and protected by proper axial rotational symmetry.~\cite{Wang2012a, Yang2014a} Such protection makes the Dirac nodes quite robust within a finite range of Hamiltonian parameters, which is exactly the reason why this class of DSM is experimentally available while the other two remain to be found. 

Despite the success in identifying Na$_3$Bi and Cd$_3$As$_2$ and the intensive studies on them, to identify more DSMs remains a big challenge. How to locate a specific material among thousands of known compounds is not clear. Here, we demonstrate a chemically intuitive approach for searching new DSMs to show the underlying physics and ideas. We choose the first DSM Na$_3$Bi as a model system for tuning the chemical degree of freedom. Three sodium ternary compounds, Na$_2$MgSn, Na$_2$MgPb, and Na$_2$CdSn, are naturally selected. Further theoretical calculations reveal that the chemical trend in the elements of the same column in periodic table plays an important role in band inversion. The proposed general design principle can be used for finding new DSMs, as well as other topological materials.



\section{Results and Discussions}

\subsection{Material design}

\begin{figure}[h]
\centering
\includegraphics[width=0.9\linewidth]{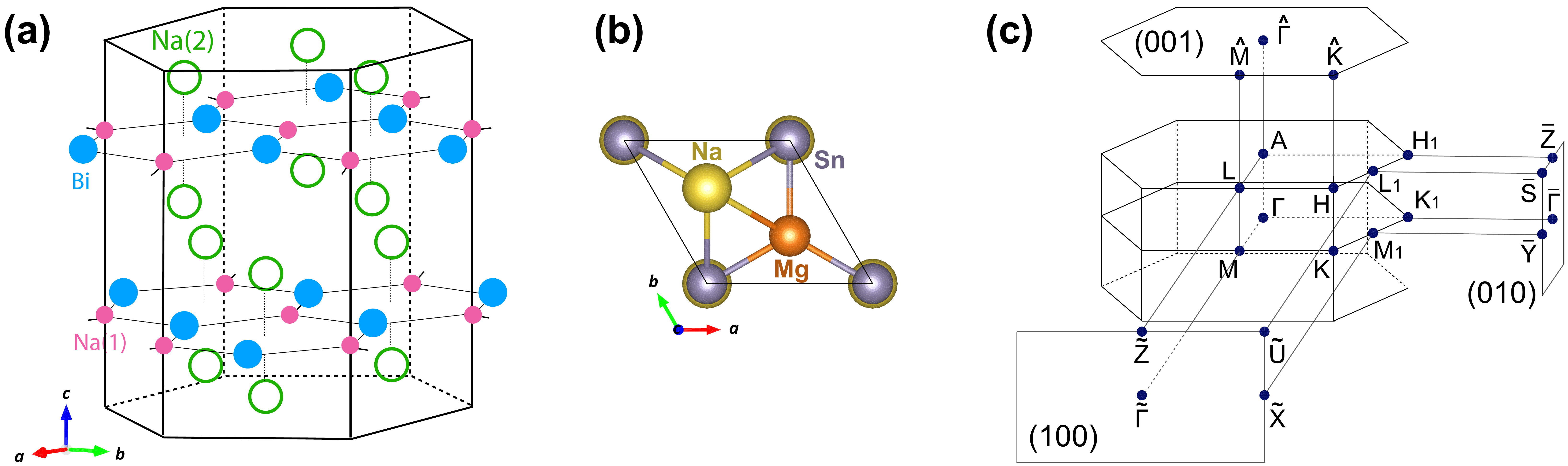}
\caption{(a) Crystal structure of Na$_3$Bi with Na(1), Na(2) and Bi sites indicated. (b) Top view of the Na$_2$MgSn unit-cell with Mg and Sn replacing Na(1) and Bi atoms in (a), respectively. (c) The bulk Brillouin zone and the projected surface Brillouin zone for (100), (010) and (001) surfaces.}
\label{lattice} 
\end{figure}

The crystal structure of Na$_3$Bi~\cite{Wang2012a,Liu2014b} can be viewed as the AB stacking of honeycomb layers along the $c$-axis, as shown in Fig.~\ref{lattice}(a). For each honeycomb layer, one Na(1) atom and one Bi atom take the A and B sub-lattice site, respectively. There are two additional Na(2) atoms above and below the Na(1)-Bi honeycomb layer to connect the Bi atoms in the neighboring layers. As a well-understood DSM, its low-energy electronic band structure has been found to be mostly determined by the Na(1) and Bi atoms in the honeycomb layer. The two crossing bands along the $\Gamma$-A direction forming Dirac nodes are dominated by Na(1)-$s$ orbitals and Bi 6$p_{x,y}$ orbitals.~\cite{Wang2012a} At $\Gamma$ point the Na(1)-$s$ bands are lower than those of Bi 6$p_{x,y}$ mainly due to two things. One is that the heavy Bi has a relatively high on-site energy for 6$p$ orbitals. The other is the interlayer coupling leads to splittings between the bonding and anti-bonding states for both $s$ and $p$ bands along $\Gamma$-A. These two crossing bands with different orbital characters have different irreducible representations along the $\Gamma$-A direction and the Dirac nodes are protected. 

Inspired by the above understanding, we notice that Na$_3$Bi can be regarded as Na$_2$Na$_1$Bi. The first two Na are on Na(2) site, which support the 3D lattice structure and also supply two electrons to the Na(1)-Bi honeycomb layer. If the crystal structure and the electronic structure could be kept the similar to those of Na$_3$Bi, one can get a new DSM material. Thus, this leads to the idea to find other potential DSMs by simply changing the atoms in the Na(1)-Bi layer. To induce band inversion, Bi should be substituted with other similar heavy metal atoms such as Pb and Sn. Since Pb and Sn have one fewer valence electron than Bi, to maintain the same band-filling, Na(1) should be substituted with atoms having two-valence electrons, such as alkaline-earth metal and II-B elements like Mg, Ca, Sr, Zn, Cd and Hg. Thus, three sodium-containing ternary compounds reported experimentally, namely Na$_2$MgSn, Na$_2$MgPb, and Na$_2$CdSn, are naturally and immediately located. Na$_2$MgSn and Na$_2$MgPb have been successfully synthesized recently~\cite{Yamada2012,Yamada2014}, while Na$_2$CdSn has been synthesized and investigated in 1980.~\cite{Matthes2014}

\begin{table}[h]
\centering
\caption{Optimized lattice constants, and lengths of the two shortest bonds (in-plane Mg/Cd-Sn/Pb bonds and vertical Na-Sn/Pb bonds) for Na$_2$MgSn, Na$_2$MgPb, and Na$_2$CdSn. The experimental data are presented in parentheses for comparison.}
\begin{tabular}{ccccccccc}
\hline
 & $a$ (\AA) & $c$ (\AA) & $d_{\rm{II-IV}}$ (\AA) & $d_{\rm{Na-IV}}$ (\AA) \\
\hline
Na$_2$MgSn & 5.078 (5.049 \cite{Yamada2012}) & 10.112 (10.095 \cite{Yamada2012}) & 2.932  (2.915 \cite{Yamada2012}) & 3.336  (3.328 \cite{Yamada2012}) \\
Na$_2$MgPb & 5.157 (5.110 \cite{Yamada2014}) & 10.240 (10.171 \cite{Yamada2014}) & 2.977 (2.950 \cite{Yamada2014}) & 3.375 (3.377 \cite{Yamada2014}) \\
Na$_2$CdSn & 5.068 (4.990 \cite{Matthes2014}) & 10.152 (10.111 \cite{Matthes2014}) & 2.926 & 3.366 \\
\hline
\end{tabular}
\label{t1}
\end{table}

Similar to Na$_3$Bi, all these compounds crystallize in hexagonal lattice with the space group $P6_3/mmc$ (\#194, $D^4_{6h}$). We take Na$_2$MgSn as an example, as demonstrated in Fig.~\ref{lattice}(b). There are four Na atoms, two Mg atoms and two Sn atoms in each unit cell. The shortest bonds are those in the Mg-Sn layer. Na and Sn atoms align along the $c$-axis connected by the second shortest bonds. The optimized lattice constants and bond lengths are listed in Table~\ref{t1}, which are in good agreements with previous experimental results. \cite{Yamada2012,Yamada2014,Matthes2014}

\begin{figure}[h]
\centering
\includegraphics[width=\linewidth]{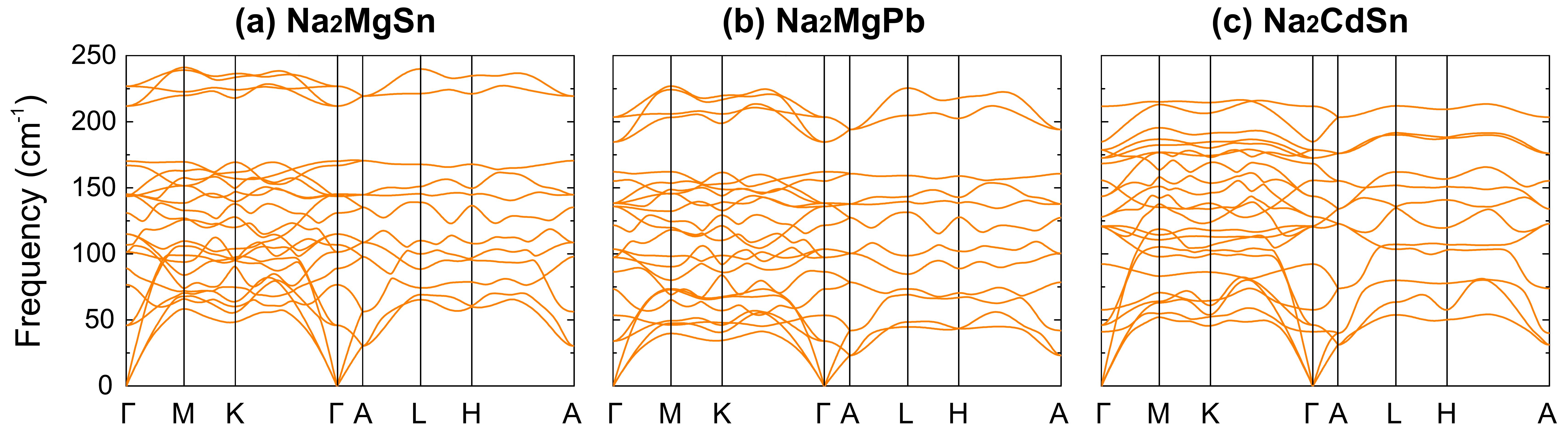}
\caption{Phonon dispersion for (a) Na$_2$MgSn, (b) Na$_2$MgPb, and (c) Na$_2$CdSn.}
\label{phonon} 
\end{figure}

For future experimental explorations, the stability of these three structures is an important aspect.~\cite{Zhang2012,Zhou2014a,Peng2017a} A material is dynamically stable when there is no imaginary phonon frequency existing in its phonon spectrum. As shown in Fig.~\ref{phonon}, no imaginary phonon frequency is found in all three materials, indicating their dynamical stability at 0 K. This is consistent with the existence of them reported by experiments. As possible candidates for DSMs, one main advantage of these sodium ternary compounds compared to Na$_3$Bi is structural dynamic stability. For Na$_3$Bi, the $P6_3/mmc$ phase has been found dynamically unstable at the ground state due to large imaginary phonon frequencies.~\cite{Cheng2014} In fact, even now the ground state of Na$_3$Bi is still under debate.~\cite{Cheng2015,Shao2017}

\subsection{Electronic structures}

\begin{figure}[h]
\centering
\includegraphics[width=\linewidth]{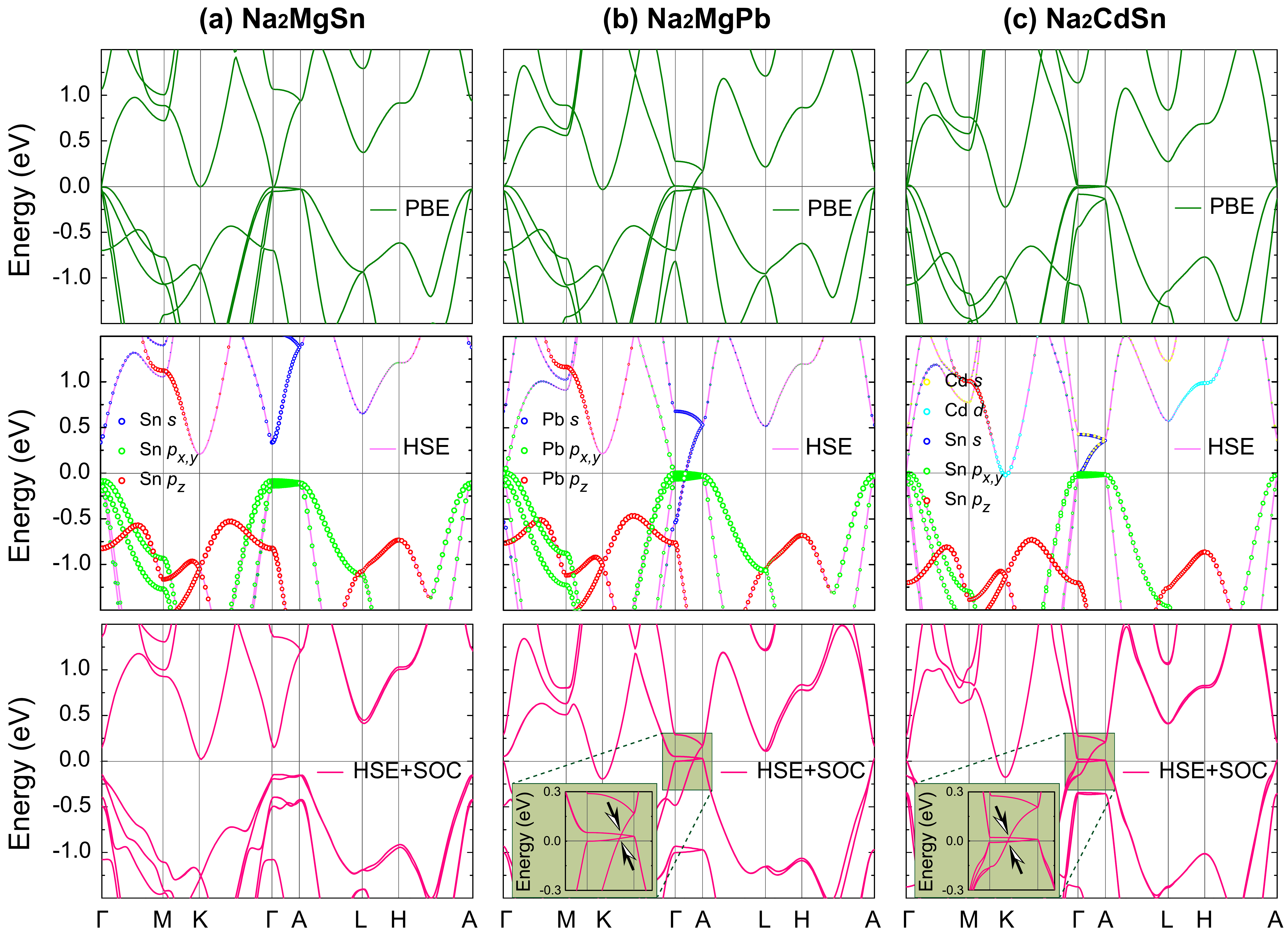}
\caption{Calculated electronic structures for (a) Na$_2$MgSn, (b) Na$_2$MgPb, and (c) Na$_2$CdSn using the PBE functional without spin-orbit coupling (top panel), and hybrid functional without (middle panel) and with (bottom panel) spin-orbit coupling. The fatted bands with the weight of atomic orbital projection near the Fermi level are present in the middle panel. The two arrows point out the two Dirac cones formed by band crossings from $s$-band and bonding, anti-bonding $p_{x,y}$-bands along $\Gamma$-A.}
\label{band} 
\end{figure}

The calculated electronic structures of all three materials using the Perdew-Burke-Ernzerhof (PBE) functional and the Heyd-Scuseria-Ernzerhof (HSE) hybrid functional are shown in the top and middle panels of Fig.~\ref{band}, respectively. The fatted bands with the weight of projected atomic orbitals are also shown in the middle panel for each of them. We focus on the band structures along $\Gamma$-A, where the band inversion and Dirac nodes happen in Na$_3$Bi.


In general, the strength of band inversion between the bands composed of $s$ orbitals (of Mg or Cd on Na(1) site) and $p$ orbitals (of Sn or Pb on Bi site) follows the order of total atomic number (mass) of the atoms in the unit cell within both PBE and HSE calculations. The overestimation of band inversion in PBE is improved by HSE calculation. One can find that the lightest Na$_2$MgSn has no band inversion and it is a normal semiconductor in HSE case. Na$_2$MgPb has the same total mass as Na$_3$Bi and is slightly lighter than the heaviest Na$_2$CdSn, but all of them have the similar band inversion along $\Gamma$-A.

The spin-orbit coupling (SOC) is further included and the band structures of them are shown in the bottom panel in Fig.~\ref{band}. Both Na$_2$MgPb and Na$_2$CdSn are DSMs with Dirac nodes on the path $\Gamma$-A, while Na$_2$MgSn is an indirect band gap of 0.13 eV. For Na$_2$MgPb and Na$_2$CdSn, one notable difference from Na$_3$Bi is that there are two pairs of Dirac nodes since the one $s$-orbital band inverts with both the bonding and anti-bonding $p_{x, y}$-orbital bands. The $s$-band belongs to $\Gamma_7$ representation while the two $p_{x,y}$ bands belong to $\Gamma_9$ representation. The splitting in the bonding and anti-bonding $p_{x, y}$ (in-plane orbitals) bands along $\Gamma$-A ($z$-direction) seems quite small, indicating the weak interlayer coupling among these in-plane orbitals along the stacking direction.

\subsection{Surface states}

\begin{figure}[h]
\centering
\includegraphics[width=\linewidth]{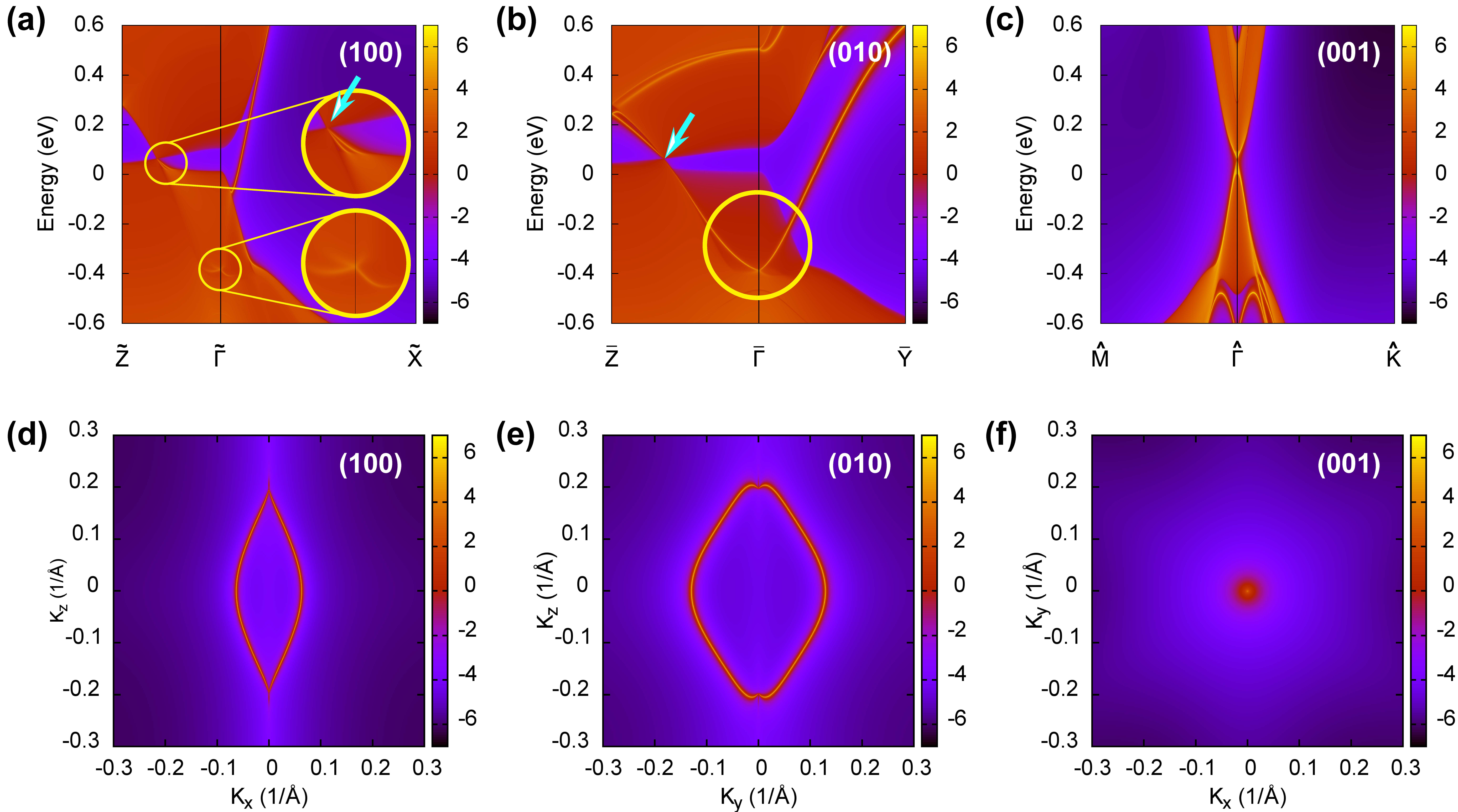}
\caption{Surface band structure for (a) (100), (b) (010), and (c) (001) surfaces of Na$_2$MgPb. The arrow points out the bulk Dirac cone, and the circle labels the topological surface states due to Z$_2$=1 in $k_z$=0 plane. The corresponding Fermi surface with Fermi level at bulk Dirac point (61 meV) is shown in (d)-(f).}
\label{density} 
\end{figure}

Similar to Na$_3$Bi, there will be surface states for DSMs Na$_2$MgPb and Na$_2$CdSn. To simulate surface states to be observed by the angle-resolved photoemission spectroscopy (ARPES), we use an iterative surface Green's function method~\cite{Zhang2010a,Wu2018}, where the HSE+SOC band structures are used in generating the maximally localized Wannier functions. The Brillouin zone of bulk and the projected surface Brillouin zones of (100), (010), and (001) planes are exactly the same as those of Na$_3$Bi,~\cite{Wang2012a} WC-type ZrTe,~\cite{Weng2016a} and KHgAs.~\cite{Wang2016k} The projected surface density of states for the (100), (010), and (001) surfaces of Na$_2$MgPb are shown in Fig.~\ref{density}(a)-(c). On both (100) and (010) side surfaces, the projection of bulk Dirac cone (pointed by the arrow) is well separated from the topological surface Dirac cone (labelled by the circle). The surface Dirac cone has its branches merging into the bulk states at the projection of 3D Dirac point, which leads to the arc like Fermi surface when the Fermi level is set at the bulk Dirac nodal point. There are two Fermi arcs touch each other at the surface projection of bulk Dirac point at 61 meV, as shown in Fig.~\ref{density}(d) and (e). For the (001) surface, the projection of bulk Dirac nodes overlaps with the surface Dirac cone as shown in Fig.~\ref{density}(c), which is similar to the case in Na$_3$Bi.~\cite{Wang2012a,Weng2016}


\begin{figure}[h]
\centering
\includegraphics[width=\linewidth]{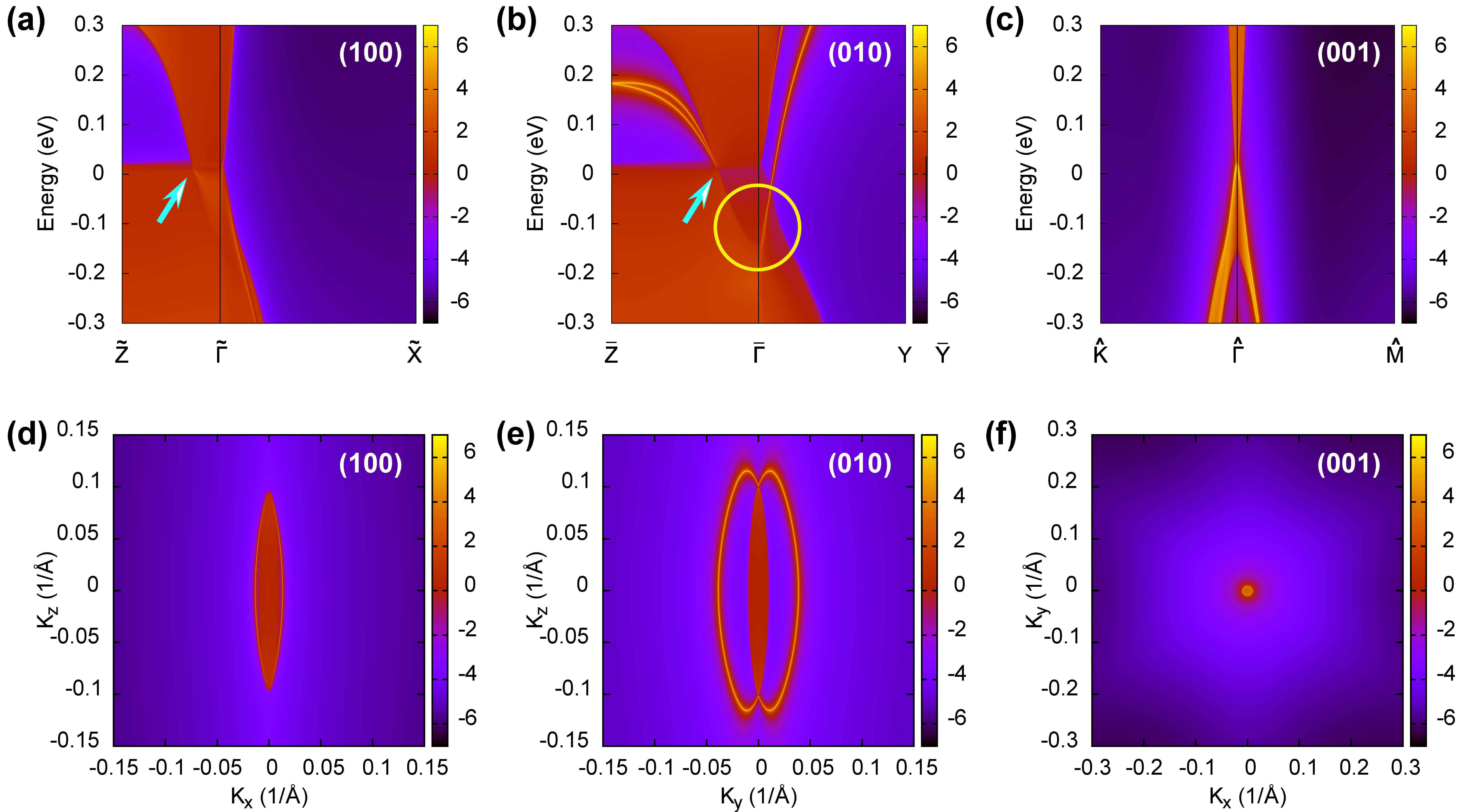}
\caption{Surface band structure for (a) (100), (b) (010), and (c) (001) surfaces of Na$_2$CdSn. The arrow points out the bulk Dirac cone, and the circle labels the topological surface states. The corresponding Fermi surface with Fermi level at bulk Dirac point (40 meV) is shown in (d)-(f).}
\label{Na2CdSn} 
\end{figure}

The projected surface density of states for the (100), (010), and (001) surfaces of Na$_2$CdSn are shown in Fig.~\ref{Na2CdSn}. For both the (100) and (010) surfaces, the bulk Dirac cone is closer to the $\Gamma$ point. Due to the smaller band splitting between the bonding and anti-bonding $p_{x, y}$-orbital bands, the nontrivial surface states of Na$_2$CdSn are not as clear as those in Na$_2$MgPb. For the (100) surface, the Fermi arcs are hidden within the projection of the bulk states on the surface. They can be well revealed in the Fermi surface plot on the (010) surface with Fermi level at bulk Dirac point of 40 meV, as shown in Fig.~\ref{Na2CdSn}(e). For the (001) surface, the surface projection of the bulk states is superposed with the nontrivial surface states, which is similar to the case in Na$_2$MgPb.


In this paper, we demonstrate an approach for searching new DSM materials by tuning the chemical degree of freedom based on material design of well-known DSM Na$_3$Bi. By keeping both the crystal and electronic structures essentially identical to Na$_3$Bi,  three compounds Na$_2$MgSn, Na$_2$MgPb, and Na$_2$CdSn are naturally located and two of them are identified as DSM candidates based on our theoretical calculations. The phonon calculations confirm that these compounds are stable than Na$_3$Bi, paving the way for experimental verification. The hybrid-functional calculations with spin-orbit coupling show that Na$_2$MgSn is an indirect band gap normal semiconductor. By substituting Sn by heavier Pb, the band inversion occurs, and the Dirac nodes due to band crossing are protected by crystal symmetry in Na$_2$MgPb. For Na$_2$CdSn, the band inversion is induced by replacing Mg with heavier Cd in Na$_2$MgSn. Moreover, the coexistence of both a bulk 3D Dirac cone and topological surface states can be observed in the projected surface density of states for side surfaces (100) and (010), which can be used as a reference for further experimental validation in ARPES or scanned tunneling microscopy measurements. We hope the idea in this example would lead to more material design efforts based on known topological materials for more successful and efficient predictions.

\section{Computational methods}

First-principles calculations are performed using the Vienna \textit{ab-initio} simulation package (VASP)~\cite{Kresse1996} based on density functional theory (DFT). The generalized gradient approximation (GGA) in the PBE parameterization for the exchange-correlation functional is used for structural relaxation. A plane-wave basis set is employed with kinetic energy cutoff of 500 eV. We use the projector-augmented-wave method and the related pseudo-potential for each element. A 11$\times$11$\times$5 \textbf{q}-mesh is used during structural relaxation for the unit cell until the energy difference is converged within 10$^{-6}$ eV, with a Hellman-Feynman force convergence threshold of 10$^{-4}$ eV/\AA. To improve the underestimation of band gap in the PBE functional, hybrid functional method based on the HSE method are adopted.~\cite{HSE1,HSE2,HSE3} The harmonic interatomic force constants (IFCs) are obtained by density functional perturbation theory using a 3$\times$3$\times$2 supercell with a 3$\times$3$\times$3 \textbf{q}-mesh. The phonon dispersion is calculated from the harmonic IFCs using the PHONOPY code.~\cite{Togo2008,Togo2015} The Wannier functions~\cite{Mostofi2014} for Cd/Mg $s$-orbital and Sn/Pb $s$-and $p$-orbitals are generated, which are used in the surface state calculations.

During the preparation of this manuscript, Ref.~\onlinecite{wan2018} proposed that Na$_2$CdSn is a topological crystalline insulator (TCI) candidate, which is consistent with our PBE+SOC calculation. From Fig.~\ref{band}(c), it is seen that both bonding and anti-bonding $s$ bands are lower than the $p_{x,y}$ bands along the whole path $\Gamma$-A. And we have confirmed that in this case it is a TCI of $Z_{12}$=8~\cite{Song2018} with mirror Chern number 2 in $m_{001}$ plane. 

\subsection*{Data Availability}

The datasets generated during and/or analysed during the current study are available from the corresponding author on reasonable request.

\section*{Acknowledgements}
The authors thank the valuable discussions with X. Wan, C. Fang, Z. Song and T. Zhang. This work is supported by the  the National Key Research and Development Program of China (No. 2016YFA0300600 and 2018YFA0305700), the National Natural Science Foundation of China (Grant Nos. 11374063 and 11674369) and the ``Strategic Priority Research Program (B)" of the Chinese Academy of Sciences (Grant No. XDB07020100).

\subsection*{Competing Interests statement}

The authors declare no competing interests.

\subsection*{Author Contributions}

H.M.W. and H.Z. designed the research. B.P., C.M.Y and H.Z. performed the calculations. B.P., H.M.W., Z.F. and H.Z. analyzed and discussed the results. B.P. and W.H.M. wrote the text of the manuscript. B.P. and C.M.Y. contributed equally to this work. All authors commented on the manuscript.


\begin{thebibliography}{43}%
\makeatletter
\providecommand \@ifxundefined [1]{%
 \@ifx{#1\undefined}
}%
\providecommand \@ifnum [1]{%
 \ifnum #1\expandafter \@firstoftwo
 \else \expandafter \@secondoftwo
 \fi
}%
\providecommand \@ifx [1]{%
 \ifx #1\expandafter \@firstoftwo
 \else \expandafter \@secondoftwo
 \fi
}%
\providecommand \natexlab [1]{#1}%
\providecommand \enquote  [1]{``#1''}%
\providecommand \bibnamefont  [1]{#1}%
\providecommand \bibfnamefont [1]{#1}%
\providecommand \citenamefont [1]{#1}%
\providecommand \href@noop [0]{\@secondoftwo}%
\providecommand \href [0]{\begingroup \@sanitize@url \@href}%
\providecommand \@href[1]{\@@startlink{#1}\@@href}%
\providecommand \@@href[1]{\endgroup#1\@@endlink}%
\providecommand \@sanitize@url [0]{\catcode `\\12\catcode `\$12\catcode
  `\&12\catcode `\#12\catcode `\^12\catcode `\_12\catcode `\%12\relax}%
\providecommand \@@startlink[1]{}%
\providecommand \@@endlink[0]{}%
\providecommand \url  [0]{\begingroup\@sanitize@url \@url }%
\providecommand \@url [1]{\endgroup\@href {#1}{\urlprefix }}%
\providecommand \urlprefix  [0]{URL }%
\providecommand \Eprint [0]{\href }%
\providecommand \doibase [0]{http://dx.doi.org/}%
\providecommand \selectlanguage [0]{\@gobble}%
\providecommand \bibinfo  [0]{\@secondoftwo}%
\providecommand \bibfield  [0]{\@secondoftwo}%
\providecommand \translation [1]{[#1]}%
\providecommand \BibitemOpen [0]{}%
\providecommand \bibitemStop [0]{}%
\providecommand \bibitemNoStop [0]{.\EOS\space}%
\providecommand \EOS [0]{\spacefactor3000\relax}%
\providecommand \BibitemShut  [1]{\csname bibitem#1\endcsname}%
\let\auto@bib@innerbib\@empty
\bibitem [{\citenamefont {Young}(2012)}]{Young2012}%
  \BibitemOpen
  \bibfield  {author} {\bibinfo {author} {\bibfnamefont {S.~M.}\
  \bibnamefont {Young \textit{et~al.}}},\ }\bibfield  {title} {\enquote {\bibinfo {title}
  {{Dirac Semimetal in Three Dimensions}},}\ }\href {\doibase
  10.1103/PhysRevLett.108.140405} {\bibfield  {journal} {\bibinfo  {journal}
  {Phys. Rev. Lett.}\ }\textbf {\bibinfo {volume} {108}},\ \bibinfo {pages}
  {140405} (\bibinfo {year} {2012})}\BibitemShut {NoStop}%
\bibitem [{\citenamefont {Wang}(2012)}]{Wang2012a}%
  \BibitemOpen
  \bibfield  {author} {\bibinfo {author} {\bibfnamefont {Zhijun}\
  \bibnamefont {Wang \textit{et~al.}}},\ }\bibfield  {title} {\enquote {\bibinfo {title}
  {{Dirac semimetal and topological phase transitions in ${A}_{3}$Bi
  ($A=\text{Na}$, K, Rb)}},}\ }\href {\doibase 10.1103/PhysRevB.85.195320}
  {\bibfield  {journal} {\bibinfo  {journal} {Phys. Rev. B}\ }\textbf {\bibinfo
  {volume} {85}},\ \bibinfo {pages} {195320} (\bibinfo {year}
  {2012})}\BibitemShut {NoStop}%
\bibitem [{\citenamefont {Wang}\ \emph {et~al.}(2013)\citenamefont {Wang},
  \citenamefont {Weng}, \citenamefont {Wu}, \citenamefont {Dai},\ and\
  \citenamefont {Fang}}]{Wang2013}%
  \BibitemOpen
  \bibfield  {author} {\bibinfo {author} {\bibfnamefont {Zhijun}\ \bibnamefont
  {Wang}}, \bibinfo {author} {\bibfnamefont {Hongming}\ \bibnamefont {Weng}},
  \bibinfo {author} {\bibfnamefont {Quansheng}\ \bibnamefont {Wu}}, \bibinfo
  {author} {\bibfnamefont {Xi}~\bibnamefont {Dai}}, \ and\ \bibinfo {author}
  {\bibfnamefont {Zhong}\ \bibnamefont {Fang}},\ }\bibfield  {title} {\enquote
  {\bibinfo {title} {{Three-dimensional Dirac semimetal and quantum transport
  in Cd${}_{3}$As${}_{2}$}},}\ }\href {\doibase 10.1103/PhysRevB.88.125427}
  {\bibfield  {journal} {\bibinfo  {journal} {Phys. Rev. B}\ }\textbf {\bibinfo
  {volume} {88}},\ \bibinfo {pages} {125427} (\bibinfo {year}
  {2013})}\BibitemShut {NoStop}%
\bibitem [{\citenamefont {Weng}\ \emph
  {et~al.}(2016{\natexlab{a}})\citenamefont {Weng}, \citenamefont {Dai},\ and\
  \citenamefont {Fang}}]{Weng2016}%
  \BibitemOpen
  \bibfield  {author} {\bibinfo {author} {\bibfnamefont {Hongming}\
  \bibnamefont {Weng}}, \bibinfo {author} {\bibfnamefont {Xi}~\bibnamefont
  {Dai}}, \ and\ \bibinfo {author} {\bibfnamefont {Zhong}\ \bibnamefont
  {Fang}},\ }\bibfield  {title} {\enquote {\bibinfo {title} {{Topological
  semimetals predicted from first-principles calculations}},}\ }\href {\doibase
  10.1088/0953-8984/28/30/303001} {\bibfield  {journal} {\bibinfo  {journal}
  {Journal of Physics: Condensed Matter}\ }\textbf {\bibinfo {volume} {28}},\
  \bibinfo {pages} {303001} (\bibinfo {year} {2016}{\natexlab{a}})}\BibitemShut
  {NoStop}%
\bibitem [{\citenamefont {Armitage}\ and\ \citenamefont
  {et~al.}(2018)}]{Armitage2018}%
  \BibitemOpen
  \bibfield  {author} {\bibinfo {author} {\bibfnamefont {N.~P.}\ \bibnamefont
  {Armitage \textit{et~al.}},}\ }\bibfield
  {title} {\enquote {\bibinfo {title} {{Weyl and Dirac semimetals in
  three-dimensional solids}},}\ }\href {\doibase 10.1103/RevModPhys.90.015001}
  {\bibfield  {journal} {\bibinfo  {journal} {Rev. Mod. Phys.}\ }\textbf
  {\bibinfo {volume} {90}},\ \bibinfo {pages} {015001} (\bibinfo {year}
  {2018})}\BibitemShut {NoStop}%
\bibitem [{\citenamefont {Novoselov}(2004)}]{Novoselov2004}%
  \BibitemOpen
  \bibfield  {author} {\bibinfo {author} {\bibfnamefont {K.~S.}\
  \bibnamefont {Novoselov \textit{et~al.}}},\ }\bibfield  {title} {\enquote {\bibinfo {title}
  {{Electric Field Effect in Atomically Thin Carbon Films}},}\ }\href {\doibase
  10.1126/science.1102896} {\bibfield  {journal} {\bibinfo  {journal}
  {Science}\ }\textbf {\bibinfo {volume} {306}},\ \bibinfo {pages} {666}
  (\bibinfo {year} {2004})}\BibitemShut {NoStop}%
\bibitem [{\citenamefont {Liu}(2014{\natexlab{a}})}]{Liu2014b}%
  \BibitemOpen
  \bibfield  {author} {\bibinfo {author} {\bibfnamefont {Z.~K.}\
  \bibnamefont {Liu \textit{et~al.}}},\ }\bibfield  {title} {\enquote {\bibinfo {title}
  {{Discovery of a Three-Dimensional Topological Dirac Semimetal, Na$_3$Bi}},}\
  }\href {\doibase 10.1126/science.1245085} {\bibfield  {journal} {\bibinfo
  {journal} {Science}\ }\textbf {\bibinfo {volume} {343}},\ \bibinfo {pages}
  {864} (\bibinfo {year} {2014}{\natexlab{a}})}\BibitemShut {NoStop}%
\bibitem [{\citenamefont {Bradlyn}(2016)}]{Bradlyn2016}%
  \BibitemOpen
  \bibfield  {author} {\bibinfo {author} {\bibfnamefont {Barry}\
  \bibnamefont {Bradlyn \textit{et~al.}}},\ }\bibfield  {title} {\enquote {\bibinfo {title}
  {{Beyond Dirac and Weyl fermions: Unconventional quasiparticles in
  conventional crystals}},}\ }\href {\doibase 10.1126/science.aaf5037}
  {\bibfield  {journal} {\bibinfo  {journal} {Science}\ }\textbf {\bibinfo
  {volume} {353}},\ \bibinfo {pages} {aaf5037} (\bibinfo {year}
  {2016})}\BibitemShut {NoStop}%
\bibitem [{\citenamefont {Bernevig}\ \emph {et~al.}(2018)\citenamefont
  {Bernevig}, \citenamefont {Weng}, \citenamefont {Fang},\ and\ \citenamefont
  {Dai}}]{Bernevig2018}%
  \BibitemOpen
  \bibfield  {author} {\bibinfo {author} {\bibfnamefont {Andrei}\ \bibnamefont
  {Bernevig}}, \bibinfo {author} {\bibfnamefont {Hongming}\ \bibnamefont
  {Weng}}, \bibinfo {author} {\bibfnamefont {Zhong}\ \bibnamefont {Fang}}, \
  and\ \bibinfo {author} {\bibfnamefont {Xi}~\bibnamefont {Dai}},\ }\bibfield
  {title} {\enquote {\bibinfo {title} {{Recent Progress in the Study of
  Topological Semimetals}},}\ }\bibfield  {booktitle} {\emph {\bibinfo
  {booktitle} {Journal of the Physical Society of Japan}},\ }\href {\doibase
  10.7566/JPSJ.87.041001} {\bibfield  {journal} {\bibinfo  {journal} {J. Phys.
  Soc. Jpn.}\ }\textbf {\bibinfo {volume} {87}},\ \bibinfo {pages} {041001}
  (\bibinfo {year} {2018})}\BibitemShut {NoStop}%
\bibitem [{\citenamefont {Orlita}(2014)}]{Orlita2014}%
  \BibitemOpen
  \bibfield  {author} {\bibinfo {author} {\bibfnamefont {M.}\
  \bibnamefont {Orlita \textit{et~al.}}},\ }\bibfield  {title} {\enquote {\bibinfo {title}
  {{Observation of three-dimensional massless Kane fermions in a zinc-blende
  crystal}},}\ }\href {\doibase 10.1038/nphys2857} {\bibfield  {journal}
  {\bibinfo  {journal} {Nat Phys}\ }\textbf {\bibinfo {volume} {10}},\ \bibinfo
  {pages} {233} (\bibinfo {year} {2014})}\BibitemShut {NoStop}%
\bibitem [{\citenamefont {Yang}\ and\ \citenamefont
  {Nagaosa}(2014)}]{Yang2014a}%
  \BibitemOpen
  \bibfield  {author} {\bibinfo {author} {\bibfnamefont {Bohm-Jung}\
  \bibnamefont {Yang}}\ and\ \bibinfo {author} {\bibfnamefont {Naoto}\
  \bibnamefont {Nagaosa}},\ }\bibfield  {title} {\enquote {\bibinfo {title}
  {{Classification of stable three-dimensional Dirac semimetals with nontrivial
  topology}},}\ }\href {\doibase 10.1038/ncomms5898} {\bibfield  {journal}
  {\bibinfo  {journal} {Nat Commun}\ }\textbf {\bibinfo {volume} {5}},\
  \bibinfo {pages} {4898} (\bibinfo {year} {2014})}\BibitemShut {NoStop}%
\bibitem [{\citenamefont {Murakami}(2007)}]{Murakami2007}%
  \BibitemOpen
  \bibfield  {author} {\bibinfo {author} {\bibfnamefont {Shuichi}\ \bibnamefont
  {Murakami}},\ }\bibfield  {title} {\enquote {\bibinfo {title} {{Phase
  transition between the quantum spin Hall and insulator phases in 3D:
  emergence of a topological gapless phase}},}\ }\href {\doibase
  10.1088/1367-2630/9/9/356} {\bibfield  {journal} {\bibinfo  {journal} {New
  Journal of Physics}\ }\textbf {\bibinfo {volume} {9}},\ \bibinfo {pages}
  {356} (\bibinfo {year} {2007})}\BibitemShut {NoStop}%
\bibitem [{\citenamefont {Weng}\ \emph {et~al.}(2017)\citenamefont {Weng},
  \citenamefont {Fang}, \citenamefont {Fang},\ and\ \citenamefont
  {Dai}}]{Weng2017}%
  \BibitemOpen
  \bibfield  {author} {\bibinfo {author} {\bibfnamefont {Hongming}\
  \bibnamefont {Weng}}, \bibinfo {author} {\bibfnamefont {Chen}\ \bibnamefont
  {Fang}}, \bibinfo {author} {\bibfnamefont {Zhong}\ \bibnamefont {Fang}}, \
  and\ \bibinfo {author} {\bibfnamefont {Xi}~\bibnamefont {Dai}},\ }\bibfield
  {title} {\enquote {\bibinfo {title} {{A new member in topological semimetals
  family}},}\ }\href {\doibase 10.1093/nsr/nwx066} {\bibfield  {journal}
  {\bibinfo  {journal} {National Science Review}\ ,\ \bibinfo {pages}
  {nwx066--nwx066}} (\bibinfo {year} {2017})}\BibitemShut {NoStop}%
\bibitem [{\citenamefont {Zdanowicz}\ and\ \citenamefont
  {Zdanowicz}(1975)}]{Zdanowicz1975}%
  \BibitemOpen
  \bibfield  {author} {\bibinfo {author} {\bibfnamefont {W}~\bibnamefont
  {Zdanowicz}}\ and\ \bibinfo {author} {\bibfnamefont {L}~\bibnamefont
  {Zdanowicz}},\ }\bibfield  {title} {\enquote {\bibinfo {title}
  {{Semiconducting Compounds of the AII BV Group}},}\ }\bibfield  {booktitle}
  {\emph {\bibinfo {booktitle} {Annual Review of Materials Science}},\ }\href
  {\doibase 10.1146/annurev.ms.05.080175.001505} {\bibfield  {journal}
  {\bibinfo  {journal} {Annu. Rev. Mater. Sci.}\ }\textbf {\bibinfo {volume}
  {5}},\ \bibinfo {pages} {301--328} (\bibinfo {year} {1975})}\BibitemShut
  {NoStop}%
\bibitem [{\citenamefont {Wan}\ \emph {et~al.}(2011)\citenamefont {Wan},
  \citenamefont {Turner}, \citenamefont {Vishwanath},\ and\ \citenamefont
  {Savrasov}}]{Wan2011}%
  \BibitemOpen
  \bibfield  {author} {\bibinfo {author} {\bibfnamefont {Xiangang}\
  \bibnamefont {Wan}}, \bibinfo {author} {\bibfnamefont {Ari~M.}\ \bibnamefont
  {Turner}}, \bibinfo {author} {\bibfnamefont {Ashvin}\ \bibnamefont
  {Vishwanath}}, \ and\ \bibinfo {author} {\bibfnamefont {Sergey~Y.}\
  \bibnamefont {Savrasov}},\ }\bibfield  {title} {\enquote {\bibinfo {title}
  {{Topological semimetal and Fermi-arc surface states in the electronic
  structure of pyrochlore iridates}},}\ }\href {\doibase
  10.1103/PhysRevB.83.205101} {\bibfield  {journal} {\bibinfo  {journal} {Phys.
  Rev. B}\ }\textbf {\bibinfo {volume} {83}},\ \bibinfo {pages} {205101}
  (\bibinfo {year} {2011})}\BibitemShut {NoStop}%
\bibitem [{\citenamefont {Xiong}(2015)}]{Xiong2015}%
  \BibitemOpen
  \bibfield  {author} {\bibinfo {author} {\bibfnamefont {Jun }\
  \bibnamefont {Xiong \textit{et~al.}}},\ }\bibfield  {title} {\enquote {\bibinfo {title}
  {{Evidence for the chiral anomaly in the Dirac semimetal Na$_3$Bi}},}\ }\href
  {\doibase 10.1126/science.aac6089} {\bibfield  {journal} {\bibinfo  {journal}
  {Science}\ }\textbf {\bibinfo {volume} {350}},\ \bibinfo {pages} {413}
  (\bibinfo {year} {2015})}\BibitemShut {NoStop}%
\bibitem [{\citenamefont {Gorbar}\ \emph {et~al.}(2014)\citenamefont {Gorbar},
  \citenamefont {Miransky},\ and\ \citenamefont {Shovkovy}}]{Gorbar2014}%
  \BibitemOpen
  \bibfield  {author} {\bibinfo {author} {\bibfnamefont {E.~V.}\ \bibnamefont
  {Gorbar}}, \bibinfo {author} {\bibfnamefont {V.~A.}\ \bibnamefont
  {Miransky}}, \ and\ \bibinfo {author} {\bibfnamefont {I.~A.}\ \bibnamefont
  {Shovkovy}},\ }\bibfield  {title} {\enquote {\bibinfo {title} {{Chiral
  anomaly, dimensional reduction, and magnetoresistivity of Weyl and Dirac
  semimetals}},}\ }\href {\doibase 10.1103/PhysRevB.89.085126} {\bibfield
  {journal} {\bibinfo  {journal} {Phys. Rev. B}\ }\textbf {\bibinfo {volume}
  {89}},\ \bibinfo {pages} {085126} (\bibinfo {year} {2014})}\BibitemShut
  {NoStop}%
\bibitem [{\citenamefont {Liu}(2014{\natexlab{b}})}]{Liu2014c}%
  \BibitemOpen
  \bibfield  {author} {\bibinfo {author} {\bibfnamefont {Z.~K.}\
  \bibnamefont {Liu \textit{et~al.}}},\ }\bibfield  {title} {\enquote {\bibinfo {title} {{A
  stable three-dimensional topological Dirac semimetal Cd$_3$As$_2$}},}\ }\href
  {\doibase 10.1038/nmat3990} {\bibfield  {journal} {\bibinfo  {journal} {Nat
  Mater}\ }\textbf {\bibinfo {volume} {13}},\ \bibinfo {pages} {677--681}
  (\bibinfo {year} {2014}{\natexlab{b}})}\BibitemShut {NoStop}%
\bibitem [{\citenamefont {Borisenko}(2014)}]{Borisenko2014}%
  \BibitemOpen
  \bibfield  {author} {\bibinfo {author} {\bibfnamefont {Sergey}\
  \bibnamefont {Borisenko \textit{et~al.}}},\ }\bibfield  {title} {\enquote {\bibinfo {title}
  {{Experimental Realization of a Three-Dimensional Dirac Semimetal}},}\ }\href
  {\doibase 10.1103/PhysRevLett.113.027603} {\bibfield  {journal} {\bibinfo
  {journal} {Phys. Rev. Lett.}\ }\textbf {\bibinfo {volume} {113}},\ \bibinfo
  {pages} {027603} (\bibinfo {year} {2014})}\BibitemShut {NoStop}%
\bibitem [{\citenamefont {Jeon}(2014)}]{Jeon2014}%
  \BibitemOpen
  \bibfield  {author} {\bibinfo {author} {\bibfnamefont {Sangjun}\
  \bibnamefont {Jeon \textit{et~al.}}},\ }\bibfield  {title} {\enquote {\bibinfo {title}
  {{Landau quantization and quasiparticle interference in the three-dimensional
  Dirac-semimetal Cd$_3$As$_2$}},}\ }\href {\doibase 10.1038/nmat4023}
  {\bibfield  {journal} {\bibinfo  {journal} {Nat Mater}\ }\textbf {\bibinfo
  {volume} {13}},\ \bibinfo {pages} {851--856} (\bibinfo {year}
  {2014})}\BibitemShut {NoStop}%
\bibitem [{\citenamefont {Neupane}(2014)}]{Neupane2014}%
  \BibitemOpen
  \bibfield  {author} {\bibinfo {author} {\bibfnamefont {Madhab}\
  \bibnamefont {Neupane \textit{et~al.}}},\ }\bibfield  {title} {\enquote {\bibinfo {title}
  {{Observation of a three-dimensional topological Dirac semimetal phase in
  high-mobility Cd$_3$As$_2$}},}\ }\href {\doibase 10.1038/ncomms4786}
  {\bibfield  {journal} {\bibinfo  {journal} {Nat Commun}\ }\textbf {\bibinfo
  {volume} {5}},\ \bibinfo {pages} {3786} (\bibinfo {year} {2014})}\BibitemShut
  {NoStop}%
\bibitem [{\citenamefont {Yamada}\ \emph {et~al.}(2012)\citenamefont {Yamada},
  \citenamefont {Deringer}, \citenamefont {Dronskowski},\ and\ \citenamefont
  {Yamane}}]{Yamada2012}%
  \BibitemOpen
  \bibfield  {author} {\bibinfo {author} {\bibfnamefont {Takahiro}\
  \bibnamefont {Yamada}}, \bibinfo {author} {\bibfnamefont {Volker~L.}\
  \bibnamefont {Deringer}}, \bibinfo {author} {\bibfnamefont {Richard}\
  \bibnamefont {Dronskowski}}, \ and\ \bibinfo {author} {\bibfnamefont
  {Hisanori}\ \bibnamefont {Yamane}},\ }\bibfield  {title} {\enquote {\bibinfo
  {title} {{Synthesis, Crystal Structure, Chemical Bonding, and Physical
  Properties of the Ternary Na/Mg Stannide Na$_2$MgSn}},}\ }\bibfield
  {booktitle} {\emph {\bibinfo {booktitle} {Inorganic Chemistry}},\ }\href
  {\doibase 10.1021/ic300184d} {\bibfield  {journal} {\bibinfo  {journal}
  {Inorg. Chem.}\ }\textbf {\bibinfo {volume} {51}},\ \bibinfo {pages}
  {4810--4816} (\bibinfo {year} {2012})}\BibitemShut {NoStop}%
\bibitem [{\citenamefont {Yamada}(2014)}]{Yamada2014}%
  \BibitemOpen
  \bibfield  {author} {\bibinfo {author} {\bibfnamefont {Takahiro}\
  \bibnamefont {Yamada \textit{et~al.}}},\ }\bibfield  {title} {\enquote {\bibinfo {title}
  {{Synthesis, Crystal Structure, and High-Temperature Phase Transition of the
  Novel Plumbide Na$_2$MgPb}},}\ }\bibfield  {booktitle} {\emph {\bibinfo
  {booktitle} {Inorganic Chemistry}},\ }\href {\doibase 10.1021/ic500466w}
  {\bibfield  {journal} {\bibinfo  {journal} {Inorg. Chem.}\ }\textbf {\bibinfo
  {volume} {53}},\ \bibinfo {pages} {5253--5259} (\bibinfo {year}
  {2014})}\BibitemShut {NoStop}%
\bibitem [{\citenamefont {Matthes}(1980)}]{Matthes2014}%
  \BibitemOpen
  \bibfield  {author} {\bibinfo {author} {\bibfnamefont {H.}~\bibnamefont
  {Matthes}, \bibfnamefont {R.~\&~Schuster}},\ }\bibfield  {title} {\enquote
  {\bibinfo {title} {{Ternary Sodium Phases with Cadmium or Mercury and Tin or
  Lead}},}\ }\href {\doibase 10.1515/znb-1980-0632} {\bibfield  {journal}
  {\bibinfo  {journal} {Zeitschrift f\"ur Naturforschung B}\ }\textbf {\bibinfo
  {volume} {35}},\ \bibinfo {pages} {778--780} (\bibinfo {year}
  {1980})}\BibitemShut {NoStop}%
\bibitem [{\citenamefont {Zhang}(2012)}]{Zhang2012}%
  \BibitemOpen
  \bibfield  {author} {\bibinfo {author} {\bibfnamefont {R.~F.}\
  \bibnamefont {Zhang \textit{et~al.}}},\ }\bibfield  {title} {\enquote {\bibinfo {title}
  {{Stability and Strength of Transition-Metal Tetraborides and Triborides}},}\
  }\href {\doibase 10.1103/PhysRevLett.108.255502} {\bibfield  {journal}
  {\bibinfo  {journal} {Phys. Rev. Lett.}\ }\textbf {\bibinfo {volume} {108}},\
  \bibinfo {pages} {255502} (\bibinfo {year} {2012})}\BibitemShut {NoStop}%
\bibitem [{\citenamefont {Zhou}(2014)}]{Zhou2014a}%
  \BibitemOpen
  \bibfield  {author} {\bibinfo {author} {\bibfnamefont {Liangcai}\
  \bibnamefont {Zhou \textit{et~al.}}},\ }\bibfield  {title} {\enquote {\bibinfo {title}
  {{Structural stability and thermodynamics of CrN magnetic phases from
  \textit{ab initio} calculations and experiment}},}\ }\href {\doibase
  10.1103/PhysRevB.90.184102} {\bibfield  {journal} {\bibinfo  {journal} {Phys.
  Rev. B}\ }\textbf {\bibinfo {volume} {90}},\ \bibinfo {pages} {184102}
  (\bibinfo {year} {2014})}\BibitemShut {NoStop}%
\bibitem [{\citenamefont {Peng}(2017)}]{Peng2017a}%
  \BibitemOpen
  \bibfield  {author} {\bibinfo {author} {\bibfnamefont {Bo}\
  \bibnamefont {Peng \textit{et~al.}}},\ }\bibfield  {title} {\enquote {\bibinfo {title}
  {{Stability and strength of atomically thin borophene from first principles
  calculations}},}\ }\bibfield  {booktitle} {\emph {\bibinfo {booktitle}
  {Materials Research Letters}},\ }\href {\doibase
  10.1080/21663831.2017.1298539} {\bibfield  {journal} {\bibinfo  {journal}
  {Materials Research Letters}\ }\textbf {\bibinfo {volume} {5}},\ \bibinfo
  {pages} {399--407} (\bibinfo {year} {2017})}\BibitemShut {NoStop}%
\bibitem [{\citenamefont {Cheng}(2014)}]{Cheng2014}%
  \BibitemOpen
  \bibfield  {author} {\bibinfo {author} {\bibfnamefont {Xiyue}\
  \bibnamefont {Cheng \textit{et~al.}}},\ }\bibfield  {title} {\enquote {\bibinfo {title}
  {{Ground-state phase in the three-dimensional topological Dirac semimetal
  Na$_3$Bi}},}\ }\href {\doibase 10.1103/PhysRevB.89.245201} {\bibfield
  {journal} {\bibinfo  {journal} {Phys. Rev. B}\ }\textbf {\bibinfo {volume}
  {89}},\ \bibinfo {pages} {245201} (\bibinfo {year} {2014})}\BibitemShut
  {NoStop}%
\bibitem [{\citenamefont {Cheng}\ \emph {et~al.}(2015)\citenamefont {Cheng},
  \citenamefont {Li}, \citenamefont {Li}, \citenamefont {Li},\ and\
  \citenamefont {Chen}}]{Cheng2015}%
  \BibitemOpen
  \bibfield  {author} {\bibinfo {author} {\bibfnamefont {Xiyue}\ \bibnamefont
  {Cheng}}, \bibinfo {author} {\bibfnamefont {Ronghan}\ \bibnamefont {Li}},
  \bibinfo {author} {\bibfnamefont {Dianzhong}\ \bibnamefont {Li}}, \bibinfo
  {author} {\bibfnamefont {Yiyi}\ \bibnamefont {Li}}, \ and\ \bibinfo {author}
  {\bibfnamefont {Xing-Qiu}\ \bibnamefont {Chen}},\ }\bibfield  {title}
  {\enquote {\bibinfo {title} {{Stable compositions and structures in the Na-Bi
  system}},}\ }\href {\doibase 10.1039/C4CP05115G} {\bibfield  {journal}
  {\bibinfo  {journal} {Phys. Chem. Chem. Phys.}\ }\textbf {\bibinfo {volume}
  {17}},\ \bibinfo {pages} {6933--6947} (\bibinfo {year} {2015})}\BibitemShut
  {NoStop}%
\bibitem [{\citenamefont {Shao}(2017)}]{Shao2017}%
  \BibitemOpen
  \bibfield  {author} {\bibinfo {author} {\bibfnamefont {Dexi}\
  \bibnamefont {Shao \textit{et~al.}}},\ }\bibfield  {title} {\enquote {\bibinfo {title}
  {Strain-induced quantum topological phase transitions in
  ${\mathrm{na}}_{3}\mathrm{Bi}$},}\ }\href {\doibase
  10.1103/PhysRevB.96.075112} {\bibfield  {journal} {\bibinfo  {journal} {Phys.
  Rev. B}\ }\textbf {\bibinfo {volume} {96}},\ \bibinfo {pages} {075112}
  (\bibinfo {year} {2017})}\BibitemShut {NoStop}%
\bibitem [{\citenamefont {Zhang}\ \emph {et~al.}(2010)\citenamefont {Zhang},
  \citenamefont {Yu}, \citenamefont {Zhang}, \citenamefont {Dai},\ and\
  \citenamefont {Fang}}]{Zhang2010a}%
  \BibitemOpen
  \bibfield  {author} {\bibinfo {author} {\bibfnamefont {Wei}\ \bibnamefont
  {Zhang}}, \bibinfo {author} {\bibfnamefont {Rui}\ \bibnamefont {Yu}},
  \bibinfo {author} {\bibfnamefont {Hai-Jun}\ \bibnamefont {Zhang}}, \bibinfo
  {author} {\bibfnamefont {Xi}~\bibnamefont {Dai}}, \ and\ \bibinfo {author}
  {\bibfnamefont {Zhong}\ \bibnamefont {Fang}},\ }\bibfield  {title} {\enquote
  {\bibinfo {title} {{First-principles studies of the three-dimensional strong
  topological insulators Bi$_2$Te$_3$, Bi$_2$Se$_3$ and Sb$_2$Te$_3$}},}\
  }\href {\doibase 10.1088/1367-2630/12/6/065013} {\bibfield  {journal}
  {\bibinfo  {journal} {New Journal of Physics}\ }\textbf {\bibinfo {volume}
  {12}},\ \bibinfo {pages} {065013} (\bibinfo {year} {2010})}\BibitemShut
  {NoStop}%
\bibitem [{\citenamefont {Wu}\ \emph {et~al.}(2018)\citenamefont {Wu},
  \citenamefont {Zhang}, \citenamefont {Song}, \citenamefont {Troyer},\ and\
  \citenamefont {Soluyanov}}]{Wu2018}%
  \BibitemOpen
  \bibfield  {author} {\bibinfo {author} {\bibfnamefont {QuanSheng}\
  \bibnamefont {Wu}}, \bibinfo {author} {\bibfnamefont {ShengNan}\ \bibnamefont
  {Zhang}}, \bibinfo {author} {\bibfnamefont {Hai-Feng}\ \bibnamefont {Song}},
  \bibinfo {author} {\bibfnamefont {Matthias}\ \bibnamefont {Troyer}}, \ and\
  \bibinfo {author} {\bibfnamefont {Alexey~A.}\ \bibnamefont {Soluyanov}},\
  }\bibfield  {title} {\enquote {\bibinfo {title} {{WannierTools: An
  open-source software package for novel topological materials}},}\ }\href
  {\doibase 10.1016/j.cpc.2017.09.033} {\bibfield  {journal} {\bibinfo
  {journal} {Computer Physics Communications}\ }\textbf {\bibinfo {volume}
  {224}},\ \bibinfo {pages} {405--416} (\bibinfo {year} {2018})}\BibitemShut
  {NoStop}%
\bibitem [{\citenamefont {Weng}\ \emph
  {et~al.}(2016{\natexlab{b}})\citenamefont {Weng}, \citenamefont {Fang},
  \citenamefont {Fang},\ and\ \citenamefont {Dai}}]{Weng2016a}%
  \BibitemOpen
  \bibfield  {author} {\bibinfo {author} {\bibfnamefont {Hongming}\
  \bibnamefont {Weng}}, \bibinfo {author} {\bibfnamefont {Chen}\ \bibnamefont
  {Fang}}, \bibinfo {author} {\bibfnamefont {Zhong}\ \bibnamefont {Fang}}, \
  and\ \bibinfo {author} {\bibfnamefont {Xi}~\bibnamefont {Dai}},\ }\bibfield
  {title} {\enquote {\bibinfo {title} {Coexistence of weyl fermion and massless
  triply degenerate nodal points},}\ }\href {\doibase
  10.1103/PhysRevB.94.165201} {\bibfield  {journal} {\bibinfo  {journal} {Phys.
  Rev. B}\ }\textbf {\bibinfo {volume} {94}},\ \bibinfo {pages} {165201}
  (\bibinfo {year} {2016}{\natexlab{b}})}\BibitemShut {NoStop}%
\bibitem [{\citenamefont {Wang}\ \emph {et~al.}(2016)\citenamefont {Wang},
  \citenamefont {Alexandradinata}, \citenamefont {Cava},\ and\ \citenamefont
  {Bernevig}}]{Wang2016k}%
  \BibitemOpen
  \bibfield  {author} {\bibinfo {author} {\bibfnamefont {Zhijun}\ \bibnamefont
  {Wang}}, \bibinfo {author} {\bibfnamefont {A.}~\bibnamefont
  {Alexandradinata}}, \bibinfo {author} {\bibfnamefont {R.~J.}\ \bibnamefont
  {Cava}}, \ and\ \bibinfo {author} {\bibfnamefont {B.~Andrei}\ \bibnamefont
  {Bernevig}},\ }\bibfield  {title} {\enquote {\bibinfo {title} {Hourglass
  fermions},}\ }\href {\doibase 10.1038/nature17410} {\bibfield  {journal}
  {\bibinfo  {journal} {Nature}\ }\textbf {\bibinfo {volume} {532}},\ \bibinfo
  {pages} {189--} (\bibinfo {year} {2016})}\BibitemShut {NoStop}%
\bibitem [{\citenamefont {Kresse}\ and\ \citenamefont
  {Furthm\"uller}(1996)}]{Kresse1996}%
  \BibitemOpen
  \bibfield  {author} {\bibinfo {author} {\bibfnamefont {G.}~\bibnamefont
  {Kresse}}\ and\ \bibinfo {author} {\bibfnamefont {J.}~\bibnamefont
  {Furthm\"uller}},\ }\bibfield  {title} {\enquote {\bibinfo {title}
  {{Efficient iterative schemes for \textit{ab initio} total-energy
  calculations using a plane-wave basis set}},}\ }\href {\doibase
  10.1103/PhysRevB.54.11169} {\bibfield  {journal} {\bibinfo  {journal} {Phys.
  Rev. B}\ }\textbf {\bibinfo {volume} {54}},\ \bibinfo {pages} {11169--11186}
  (\bibinfo {year} {1996})}\BibitemShut {NoStop}%
\bibitem [{\citenamefont {Heyd}\ \emph {et~al.}(2003)\citenamefont {Heyd},
  \citenamefont {Scuseria},\ and\ \citenamefont {Ernzerhof}}]{HSE1}%
  \BibitemOpen
  \bibfield  {author} {\bibinfo {author} {\bibfnamefont {Jochen}\ \bibnamefont
  {Heyd}}, \bibinfo {author} {\bibfnamefont {Gustavo~E.}\ \bibnamefont
  {Scuseria}}, \ and\ \bibinfo {author} {\bibfnamefont {Matthias}\ \bibnamefont
  {Ernzerhof}},\ }\bibfield  {title} {\enquote {\bibinfo {title} {{Hybrid
  functionals based on a screened Coulomb potential}},}\ }\href {\doibase
  10.1063/1.1564060} {\bibfield  {journal} {\bibinfo  {journal} {J. Chem.
  Phys.}\ }\textbf {\bibinfo {volume} {118}},\ \bibinfo {pages} {8207}
  (\bibinfo {year} {2003})}\BibitemShut {NoStop}%
\bibitem [{\citenamefont {Heyd}\ \emph {et~al.}(2006)\citenamefont {Heyd},
  \citenamefont {Scuseria},\ and\ \citenamefont {Ernzerhof}}]{HSE2}%
  \BibitemOpen
  \bibfield  {author} {\bibinfo {author} {\bibfnamefont {Jochen}\ \bibnamefont
  {Heyd}}, \bibinfo {author} {\bibfnamefont {Gustavo~E.}\ \bibnamefont
  {Scuseria}}, \ and\ \bibinfo {author} {\bibfnamefont {Matthias}\ \bibnamefont
  {Ernzerhof}},\ }\bibfield  {title} {\enquote {\bibinfo {title} {{Erratum:
  鈥淗ybrid functionals based on a screened Coulomb potential鈥?[J. Chem.
  Phys.118, 8207 (2003)]}},}\ }\href {\doibase 10.1063/1.2204597} {\bibfield
  {journal} {\bibinfo  {journal} {J. Chem. Phys.}\ }\textbf {\bibinfo {volume}
  {124}},\ \bibinfo {pages} {219906} (\bibinfo {year} {2006})}\BibitemShut
  {NoStop}%
\bibitem [{\citenamefont {Peralta}\ \emph {et~al.}(2006)\citenamefont
  {Peralta}, \citenamefont {Heyd}, \citenamefont {Scuseria},\ and\
  \citenamefont {Martin}}]{HSE3}%
  \BibitemOpen
  \bibfield  {author} {\bibinfo {author} {\bibfnamefont {Juan~E.}\ \bibnamefont
  {Peralta}}, \bibinfo {author} {\bibfnamefont {Jochen}\ \bibnamefont {Heyd}},
  \bibinfo {author} {\bibfnamefont {Gustavo~E.}\ \bibnamefont {Scuseria}}, \
  and\ \bibinfo {author} {\bibfnamefont {Richard~L.}\ \bibnamefont {Martin}},\
  }\bibfield  {title} {\enquote {\bibinfo {title} {{Spin-orbit splittings and
  energy band gaps calculated with the Heyd-Scuseria-Ernzerhof screened hybrid
  functional}},}\ }\href {\doibase 10.1103/PhysRevB.74.073101} {\bibfield
  {journal} {\bibinfo  {journal} {Phys. Rev. B}\ }\textbf {\bibinfo {volume}
  {74}},\ \bibinfo {pages} {073101} (\bibinfo {year} {2006})}\BibitemShut
  {NoStop}%
\bibitem [{\citenamefont {Togo}\ \emph {et~al.}(2008)\citenamefont {Togo},
  \citenamefont {Oba},\ and\ \citenamefont {Tanaka}}]{Togo2008}%
  \BibitemOpen
  \bibfield  {author} {\bibinfo {author} {\bibfnamefont {Atsushi}\ \bibnamefont
  {Togo}}, \bibinfo {author} {\bibfnamefont {Fumiyasu}\ \bibnamefont {Oba}}, \
  and\ \bibinfo {author} {\bibfnamefont {Isao}\ \bibnamefont {Tanaka}},\
  }\bibfield  {title} {\enquote {\bibinfo {title} {{First-principles
  calculations of the ferroelastic transition between rutile-type and
  CaCl$_{2}$-type SiO$_{2}$ at high pressures}},}\ }\href {\doibase
  10.1103/PhysRevB.78.134106} {\bibfield  {journal} {\bibinfo  {journal} {Phys.
  Rev. B}\ }\textbf {\bibinfo {volume} {78}},\ \bibinfo {pages} {134106}
  (\bibinfo {year} {2008})}\BibitemShut {NoStop}%
\bibitem [{\citenamefont {Togo}\ and\ \citenamefont {Tanaka}(2015)}]{Togo2015}%
  \BibitemOpen
  \bibfield  {author} {\bibinfo {author} {\bibfnamefont {Atsushi}\ \bibnamefont
  {Togo}}\ and\ \bibinfo {author} {\bibfnamefont {Isao}\ \bibnamefont
  {Tanaka}},\ }\bibfield  {title} {\enquote {\bibinfo {title} {{First
  principles phonon calculations in materials science}},}\ }\href {\doibase
  http://dx.doi.org/10.1016/j.scriptamat.2015.07.021} {\bibfield  {journal}
  {\bibinfo  {journal} {Scripta Materialia}\ }\textbf {\bibinfo {volume}
  {108}},\ \bibinfo {pages} {1--5} (\bibinfo {year} {2015})}\BibitemShut
  {NoStop}%
\bibitem [{\citenamefont {Mostofi}(2014)}]{Mostofi2014}%
  \BibitemOpen
  \bibfield  {author} {\bibinfo {author} {\bibfnamefont {Arash A.}\
  \bibnamefont {Mostofi \textit{et~al.}}},\ }\bibfield  {title} {\enquote {\bibinfo {title}
  {{An updated version of Wannier90: A tool for obtaining maximally-localised
  Wannier functions}},}\ }\href {\doibase 10.1016/j.cpc.2014.05.003} {\bibfield
   {journal} {\bibinfo  {journal} {Computer Physics Communications}\ }\textbf
  {\bibinfo {volume} {185}},\ \bibinfo {pages} {2309--2310} (\bibinfo {year}
  {2014})}\BibitemShut {NoStop}%
\bibitem [{\citenamefont {Tang}\ \emph {et~al.}(2018)\citenamefont {Tang},
  \citenamefont {Po}, \citenamefont {Vishwanath},\ and\ \citenamefont
  {Wan}}]{wan2018}%
  \BibitemOpen
  \bibfield  {author} {\bibinfo {author} {\bibfnamefont {F.}~\bibnamefont
  {Tang}}, \bibinfo {author} {\bibfnamefont {H.~C.}\ \bibnamefont {Po}},
  \bibinfo {author} {\bibfnamefont {A.}~\bibnamefont {Vishwanath}}, \ and\
  \bibinfo {author} {\bibfnamefont {X.}~\bibnamefont {Wan}},\ }\bibfield
  {title} {\enquote {\bibinfo {title} {{Topological Materials Discovery By
  Large-order symmetry indicators}},}\ }\href@noop {} {\bibfield  {journal}
  {\bibinfo  {journal} {arXiv}\ }\textbf {\bibinfo {volume} {1806}},\ \bibinfo
  {pages} {04128} (\bibinfo {year} {2018})}\BibitemShut {NoStop}%
\bibitem [{\citenamefont {Song}\ \emph {et~al.}(2018)\citenamefont {Song},
  \citenamefont {Zhang}, \citenamefont {Fang},\ and\ \citenamefont
  {Fang}}]{Song2018}%
  \BibitemOpen
  \bibfield  {author} {\bibinfo {author} {\bibfnamefont {Zhida}\ \bibnamefont
  {Song}}, \bibinfo {author} {\bibfnamefont {Tiantian}\ \bibnamefont {Zhang}},
  \bibinfo {author} {\bibfnamefont {Zhong}\ \bibnamefont {Fang}}, \ and\
  \bibinfo {author} {\bibfnamefont {Chen}\ \bibnamefont {Fang}},\ }\bibfield
  {title} {\enquote {\bibinfo {title} {Quantitative mappings between symmetry
  and topology in solids},}\ }\href {\doibase 10.1038/s41467-018-06010-w}
  {\bibfield  {journal} {\bibinfo  {journal} {Nature Communications}\ }\textbf
  {\bibinfo {volume} {9}},\ \bibinfo {pages} {3530--} (\bibinfo {year}
  {2018})}\BibitemShut {NoStop}%
\end{thebibliography}

%

\end{document}